\documentclass[lineno]{JFM-FLM_Au}
\usepackage{graphicx,cancel}
\usepackage{float}
\usepackage{caption}
\usepackage{newtxtext}
\usepackage{newtxmath}
\usepackage{natbib}
\usepackage{hyperref,amsmath,amssymb,subfig,subcaption}

\lefttitle{Shambhu Anil, Pushpavanam Subramaniam}
\righttitle{Journal of Fluid Mechanics}

\title{Resonance frequency of different interfacial modes of a slug trapped in a milli-channel}

\author{Shambhu Anil\aff{1},
	\and Pushpavanam Subramaniam\aff{1}}

\affiliation{\aff{1}Department of Chemical Engineering, Indian Institute of Technology, Madras, Chennai, 600036, Tamil Nadu, India}

\corresau{Pushpavanam Subramaniam, \email{spush@iitm.ac.in}}

\begin{document}
	\maketitle
	
	\begin{abstract}
		Several Lab – on – chip applications such as cell lysis, micromixing and micropumping are based on flows induced by acoustically excited oscillatory bubbles. For high efficiency, the system must be operated at its resonant frequency. In most systems the bubbles are present in a confined geometry where the resonant frequency is determined by the nature of confinement and its coupling with liquid flow.  In this work, we determine the resonant frequencies corresponding to surface modes of oscillation of a rectangular gas slug confined at one end of a milli -channel using perturbation techniques and matched asymptotic expansions. These are verified using simulations in Ansys Fluent and experimental results in similar geometry.
	\end{abstract}
	
	\section{Introduction}
	\label{sec:Introduction}
	Numerous Lab – On – Chip operations based on acoustically excited microbubbles have been recently developed (\cite{Hashmi2012}). The experimental setup typically consists of a microchannel fabricated from a soft material, like Poly Dimethyl Siloxane (PDMS), attached to a glass substrate on which a piezoelectric transducer is embedded. On activating this transducer, the localised pressure fluctuations induce oscillation of the gas - liquid interface. The steady streaming caused by such oscillating microbubbles is called Cavitation Microstreaming. (\cite{Davidson1971})\\
	
	\cite{Tho2007} quantified the flow field induced by single and multiple bubbles. They showed that different surface modes were observed at specific driving frequencies, and these affected the streaming patterns and velocity. \cite{Wang2013} experimentally found the amplitudes of different modes on the surface of a sessile semi – cylindrical microbubble. They also compared it with a semi – analytical solution found using perturbation theory and matched asymptotic expansions. Here the gas bubble is lodged in the side channel of a microfluidic system which on excitation, forces the fluid surrounding it to circulate around the bubble. Such systems have applications in particle sorting and particle trapping (\cite{Thameem2016}). \cite{Wang2013} also found that the number of surface modes on the bubble increases with increase in frequency.\\
	
	\cite{Rallabandi2013} established a theoretical framework for the experimental system investigated by \cite{Wang2013}. This semi - analytical study showed that the interaction of different surface modes is responsible for the fountain like steady flow from the bubble. Boundary layer effects reversed this fountain at higher frequencies.\\
	Gas slugs are used in drug delivery applications in various Lab-on-chip devices (\cite{Etminan2021}). These in conjunction with acoustic streaming have been exploited for active particle separation and micro-mixing (\cite{SinemOrbayAdemOzcelikJamesLataMuratKaynakMengxiWu2019}). The streaming velocity is maximum when the system is operated at its resonant frequency. The resonant frequency of a bubble in an unbounded medium, called Minnaert frequency (\cite{Minnaert1933}), has been studied extensively. In this study we determine the resonant frequency of a gas slug trapped at one end of a millichannel as a function of slug size, physical properties of fluids, amplitude and frequency of driving far field pressure oscillation. The effect of confinement by channel walls/geometry is also considered.\\
	
	Our objective here is to develop an analytical framework based on perturbation techniques and matched asymptotic expansion to predict the resonant frequencies of different surface modes that appear on a trapped gas slug oscillating in response to an external acoustic field. The analytical formulation used in this study is like that adopted by \cite{Wang2013}. The approach considers the confinement effect of the walls and the coupling of the gas and liquid phases across the interface. We compare our theoretical predictions of the resonant frequencies of different surface modes with numerical simulations done in ANSYS Fluent. \\
	
	The theoretical model, characteristic scales and governing equations are discussed in $\S$ 2. $\S$ 3 details the method of solution to obtain the oscillatory solution at leading order. $\S$ 4 describes the numerical simulations using ANSYS Fluent for validating the theoretical findings. Finally, $\S$ 5 discusses the key findings and inferences derived from this study.
	\section{Problem formulation}
	\begin{figure}
		\centering
		\includegraphics[width=0.7\linewidth]{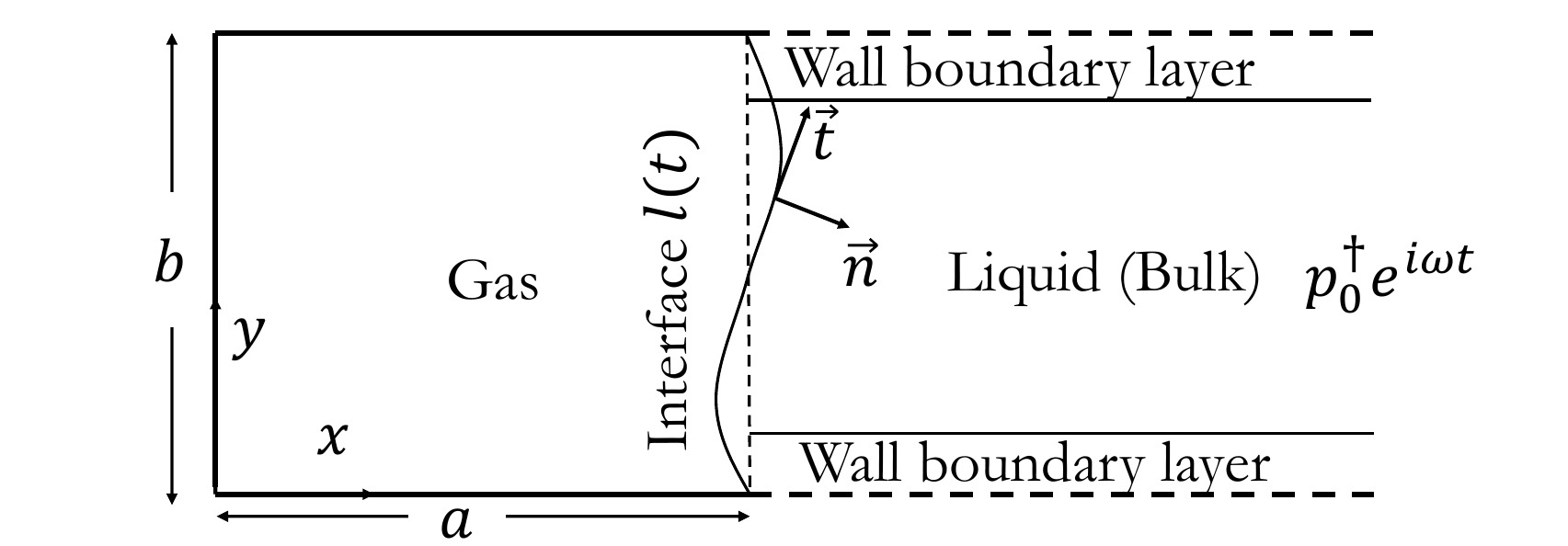}
		\caption{A schematic of the system under consideration showing the interface at the base state and the deformed state. The far field oscillatory pressure shown on the right-hand side is applied only for the numerical simulations. Dashed and solid line at the walls indicate super hydrophilic and super hydrophobic sections of the channel respectively.}
		\label{Figure 1}
	\end{figure}
	We model a trapped air slug in a water – filled millichannel (Figure \ref{Figure 1}). We assume the gas liquid interface is pinned at the walls and is flat at base state. This can be achieved experimentally by having the milli-channel wall super hydrophilic where liquid is present and super hydrophobic where air is present. A flat interface can also be attained by embedding groves that slant facing the liquid phase. (\cite{Chang2022})
	\subsection{Governing equation and characteristic scales}
	We restrict the analysis to cartesian coordinates ($x$ and $y$) and assume the system extends to infinity in the $z$ direction. The dimensionless continuity and Navier – Stokes equations are given as:
	\begin{flalign}
		\frac{\partial u}{\partial x}+\frac{\partial v}{\partial y} &=0\label{conti}\\
		\frac{\partial u}{\partial t}+\varepsilon \mathbf{u}\cdot\nabla u&=-\frac{\partial p}{\partial x}+\frac{1}{Re}\bigg(\frac{\partial^2}{\partial x^2}+k^2\frac{\partial^2}{\partial y^2}\bigg)u\label{navstokesx}\\
		\frac{\partial v}{\partial t}+\varepsilon \mathbf{u}\cdot\nabla v&=-k^2\frac{\partial p}{\partial y}+\frac{1}{Re}\bigg(\frac{\partial^2}{\partial x^2}+k^2\frac{\partial^2}{\partial y^2}\bigg)v\label{navstokesy}
	\end{flalign}
	Here $p$, $u$, $v$ are the dimensionless pressure, velocity components along $x$ and $y$ directions respectively in the liquid. The chosen characteristic length scales are, $x_{ch}= a$ and $y_{ch}=b$ and $k$ is the aspect ratio of the rectangular gas slug $(k=a/b)$ We choose the time scale as $t_{ch}=1/\omega$  where $\omega$ is the angular frequency of the oscillation of the flow field. $Re$ is the Reynolds number defined as $(\omega a^2)/\nu$. $p_{ch}=\varepsilon\rho\omega^2 a^2$ is the characteristic pressure. The characteristic horizontal streaming velocity is $u_{ch}=\varepsilon\omega a$. From continuity equation \ref{conti}, we obtain the vertical characteristic velocity as, $v_{ch}=\varepsilon\omega b$.\\
	
	Cavitation microstreaming occurs at large  Strouhal numbers (the ratio of the driving flow to observed flow) $St=\omega a/u_{ch}$. From the definition of $u_{ch}$, we see that the inverse of Strouhal number is $\varepsilon$ which is an infinitesimal quantity at large Strouhal numbers. We also specify the interface oscillation explicitly as 
	\begin{equation}
		x=l(y,t)=1-\varepsilon i h(y)e^{it}\label{interface}
	\end{equation}
	Here, $h(y)$ is the yet to be determined interface deformation. The amplitude of the interface deformation is $\varepsilon a$, which when divided by $t_{ch}=1/\omega$ gives us the corresponding $u_{ch}$. In this work, the deformation is assumed to be much lower than the length of the slug $(\varepsilon<<1)$.This enables us to analyse the system behaviour using perturbation techniques with $\varepsilon$ as the perturbation parameter.\\
	
	We also assume that the gas is inviscid, compressible and undergoes polytropic expansion and contraction with $\Gamma$ as the polytropic coefficient. The liquid is assumed to be Newtonian and incompressible. The pinning of the interface implies $h(0)=h(1)=0$. These assumptions allow us to analyse the system using rectangular cartesian coordinates, while retaining the important physics.\\
	
	The governing equations \ref{conti}, \ref{navstokesx} and \ref{navstokesy} are subject to no slip and no penetration boundary condition at the walls. At the gas liquid interface, we impose zero shear stress, normal stress balance and kinematic boundary condition.
	\section{Method of solution}
	Our objective is to determine the resonance frequency, considering the viscous and surface tension effects in the system. For this, we seek solutions to velocity and pressure in the form of a perturbation series in $\varepsilon$.
	\begin{flalign}
		u=u_0 + \varepsilon u_1+O(\varepsilon^2)\label{pertu}\\
		v=v_0 + \varepsilon v_1+O(\varepsilon^2\label{pertv})\\
		p=p_0 + \varepsilon p_1+O(\varepsilon^2)\label{pertp}
	\end{flalign}
	In  equations \ref{navstokesx} and \ref{navstokesy}, $1/Re$ is also a small parameter defined as  $1/Re=\delta^2/2$, where $\delta$ is the Stokes layer thickness given by $\sqrt{\frac{2\nu}{\omega a^2}}$. For cavitation microstreaming, the driving frequencies $(\omega)$ are high, implying $\delta<<1$. We focus on the case where our system undergoes Rayleigh – Nyborg – Westerwelt streaming for which $\varepsilon<<\delta<<1$.This assumption lets us use a regular perturbation expansion in $\varepsilon$.\\
	
	In acoustically driven flows with confinement, there arises a slip at the walls. This is caused by boundary layer streaming or Schlichting streaming, arising from viscous dissipation in boundary layers. The separation of length scales as $\delta<<1$ enables us to use matched asymptotic expansions to account for this slip arising from the boundary layer. We divide the domain into two regions: (i) an Inner region or Wall boundary layer and (ii) Outer region or Bulk as shown in Figure \ref{Figure 1}. The leading order oscillatory velocity field (identified by subscript 0) in the wall boundary layer and bulk region are matched to obtain a composite solution throughout the domain. 
	\subsection{Leading order governing equations}
	Since the flow field is two dimensional, we seek the solution in terms of stream function. 
	\begin{flalign}
		u_0&=\frac{\partial \psi_0}{\partial y}\label{streamu}\\
		v_0&=-\frac{\partial \psi_0}{\partial x}\label{streamv}
	\end{flalign}
	Substituting equations \ref{streamu} and \ref{streamv} in Navier - Stokes Equation and collecting the leading order terms from the governing equation gives,
	\begin{flalign}
		\frac{\partial}{\partial t}\bigg(\frac{\partial^2}{\partial x^2}+k^2\frac{\partial^2}{\partial y^2}\bigg)\psi_0=\frac{1}{Re}\bigg(\frac{\partial^2}{\partial x^2}+k^2\frac{\partial^2}{\partial y^2}\bigg)^2\psi_0\label{biharmonic0}
	\end{flalign}
	The boundary conditions at the interface are defined for $x=l(y,t)$. We use domain perturbation around the steady flat interface to obtain the boundary conditions at $x=1$. This gives the dimensionless boundary conditions:
	\begin{flalign}
		\text{Kinematic boundary condition at interface: }\frac{\partial\psi_0}{\partial y}\bigg|_{x=1}&=h(y)e^{it}\label{bdry:kinematic}\\
		\text{Tangential stress balance at interface: }\frac{\partial^2\psi_0}{\partial x^2}-k^2\frac{\partial^2\psi_0}{\partial y^2}\bigg|_{x=1}&=0\label{bdry:tangential}\\
		\text{Normal stress balance at the interface: }p_i^{(0)}-p_0+\frac{1}{Re}\tau_{xx}^{(0)}-\frac{1}{\varepsilon We}\kappa\bigg|_{x=1}&=0\\
		\text{No slip and no penetration at the walls: }\frac{\partial\psi_0}{\partial y}\bigg|_{y=0,1}=-\frac{\partial\psi_0}{\partial x}\bigg|_{y=0,1}&=0\label{bdry:noslipnopen}
	\end{flalign}
	$p_i^{(0)}$ and $p_0$ are the pressure in the gas phase and liquid phase respectively. $\tau_{xx}$ and $\kappa$ are the normal viscous stress and curvature of the interface. Also, $We$ is the Weber number defined as ratio of inertial to surface tension forces, $We = \frac{\rho\omega^2a^3}{\gamma}$
	\subsubsection{Outer region or bulk solution}
	The outer or bulk solution is valid in the region away from walls, where inertial effects are significant. We seek solutions for the modified biharmonic equation \ref{biharmonic0}, which are periodic in $y$ and $t$ as: $\psi_0=f(x)e^{i\lambda y}e^{it}$. Here $f(x)$ is as follows
	\begin{flalign}
		f(x)=C_1e^{\xi x}+C_2e^{-\xi x}+C_3e^{-k\lambda x}+C_4e^{k\lambda x}
	\end{flalign}
	Where $\xi = \sqrt{iRe+k^2\lambda^2}$. At high Reynolds numbers, $\frac{k^2\lambda^2}{Re}$ becomes very small. The real part of $\xi$, obtained using a Taylor series expansion in $\frac{k^2\lambda^2}{Re}$ $=\frac{k^2\lambda^2\delta^2}{2}$ is $\mathbb{R}(\xi)=\frac{1}{\delta}+\frac{1}{4}k^2\lambda^2\delta+O(\delta^3)$ which is positive. Since the flow should be bounded as $x\rightarrow\infty$, the constants $C_1$ and $C_4$ must be zero. Hence,
	\begin{flalign}
		f(x)=C_2e^{-\xi x}+C_3e^{-k\lambda x}
	\end{flalign}
	The no penetration boundary condition \ref{bdry:noslipnopen} at the two walls yield $\lambda=n\pi$, for $n=0,1,2,3$. This gives,
	\begin{flalign}
		\psi_0=\sum_{n=0}^{\infty}(C_2e^{-\xi_nx}+C_3e^{-kn\pi x})\sin{n\pi y}e^{it}
	\end{flalign}
	Here, $\xi_n=\sqrt{iRe+k^2n^2\pi^2}$. The kinematic and tangential stress boundary conditions (\ref{bdry:kinematic}, \ref{bdry:tangential}) yield,
	\begin{flalign}
		C_2=\sum_{n=0}^{\infty}A_ni\delta^2k^2n\pi e^{\xi_n}\label{sol:C2}\\
		C_3=\sum_{n=0}^{\infty}A_n\bigg(\frac{1}{n\pi}-i\delta^2k^2n\pi\bigg)e^{kn\pi}\label{sol:C3}
	\end{flalign}
	This gives,
	\begin{equation}
		\psi_0=\sum_{n=0}^{\infty}A_n\bigg(\frac{1}{n\pi}e^{-kn\pi s}+i\delta^2k^2n\pi(e^{-\xi_ns}-e^{-kn\pi s})\bigg)\sin{n\pi y}e^{it}
	\end{equation}
	Here, $s=x-1$ and $A_n$ is the amplitude of the n$^{\text{th}}$ surface mode of the interface. In this outer region, the no slip boundary condition is not imposed. Instead the horizontal slip generated from the bulk is matched with the wall boundary layer.
	\subsubsection{Inner region or Wall boundary layer solution}
	In the wall boundary layer, the $y$ variable is scaled as $\eta=\frac{y}{\delta}$ and $\eta=\frac{1-y}{\delta}$ near the bottom $(y=0)$ and top $(y=1)$ walls respectively. The stream function $\psi_0$ is also scaled as $\psi_{bl}=\frac{\psi_0}{\delta}$. Where, $\psi_{bl}$ is the stream function in the boundary layer. This scaling allows the flow in a small region close to the walls to be analyzed. This inner domain spans from $\eta=0$ at the walls to $\eta=\infty$ as $\delta<<1$. The rescaled stream function is governed by:
	\begin{flalign}
		\frac{\partial}{\partial t}\bigg(\frac{\partial^2}{\partial x^2}+\frac{k^2}{\delta^2}\frac{\partial^2}{\partial\eta^2}\bigg)\psi_{bl}=\frac{1}{Re}\bigg(\frac{\partial^2}{\partial x^2}+\frac{k^2}{\delta^2}\frac{\partial^2}{\partial\eta^2}\bigg)^2\psi_{bl}\label{eqn:wblinitial}
	\end{flalign}
	We seek $\psi_{bl}$ as $\psi_{bl}=g(x,\eta)e^{it}$ and impose the no slip and no penetration boundary conditions at the walls. As $\delta<<1$, the governing equation (\ref{eqn:wblinitial}) reduces to,
	\begin{flalign}
		\frac{2i}{k^2}\frac{\partial^2 g}{\partial \eta^2}=\frac{\partial^4 g}{\partial \eta^4}+O(\delta^2)
	\end{flalign}
	This is subject to boundary conditions as follows,
	\begin{flalign}
		\text{No penetration at walls: }\frac{\partial \psi_{bl}}{\partial x}\bigg|_{\eta=0}&=0\label{bdry:wbl_nopen}\\
		\text{No slip at walls: }\frac{\partial\psi_{bl}}{\partial \eta}\bigg|_{\eta=0}&=0\label{bdry:wbl_noslip}\\
		\text{Far field condition: }\frac{\partial\psi_{bl}}{\partial \eta}\bigg|_{\eta\rightarrow\infty} &\text{is bounded}\label{bdry:farfield}
	\end{flalign}
	This yields the expression for $\psi_{bl}$ as,
	\begin{flalign}
		\psi_{bl}=C(x)\bigg(\eta-\frac{1-i}{2}k\big(1-e^{-\frac{(1+i)\eta}{k}}\big)\bigg)e^{it}
	\end{flalign}
	More details of the derivation maybe found in Supplementary section 1. \cite{Riley2001} analysed the wall boundary layer flow field in the presence of a far field (Bulk) oscillatory flow. Our oscillatory boundary layer solution reduces to that of Riley’s when the aspect ratio, $k=1$. Our system can be used to experimentally realize the boundary layer streaming described by Riley and $k$ can be used to control this flow. To match the dominant horizontal velocity at the edge of wall boundary layer with the bulk, we use Prandtl matching (\cite{Fraenkel1969}) and impose
	\begin{flalign}
		\frac{\partial\psi_{bl}}{\partial\eta}\bigg|_{\eta\rightarrow\infty}=\frac{\partial\psi_{0}}{\partial y}\bigg|_{y=0}\text{ for the lower and }\frac{\partial\psi_{bl}}{\partial\eta}\bigg|_{\eta\rightarrow\infty}=\frac{\partial\psi_{0}}{\partial y}\bigg|_{y=1}\text{ for the upper wall}
	\end{flalign}
	This helps us determine $C(x)$ and gives the leading order stream function in the wall boundary layer,
	\begin{flalign}
		\psi_{bl}=\sum_{n=0}^{\infty}A_n\bigg(e^{-kn\pi s}+i\delta^2k^2n^2\pi^2(e^{-\xi_ns}-e^{-kn\pi s})\bigg)\bigg(\eta-\frac{1-i}{2}k\big(1-e^{-\frac{(1+i)\eta}{k}}\big)\bigg)e^{it}
	\end{flalign} 
	\subsubsection{Composite leading order solution}
	The composite velocity field in the entire domain is the ‘union’ of velocity field between bulk and boundary layers. Since the velocity is continuous and purely horizontal at the edge of boundary layers, we subtract the overlapping horizontal velocity to avoid double counting. This yields,
	\begin{flalign}
		u_{comp}&=u_{bulk}+u_{bl}-\big(u_{bulk}\big|_{y=0}+u_{bulk}\big|_{y=1}\big)
	\end{flalign} 
	\begin{multline}
		u_{comp}=\sum_{n=0}^{\infty}A_n\big(e^{-kn\pi s}+i\delta^2k^2n^2\pi^2(e^{-\xi_ns}-e^{-kn\pi s})\big)\\
		\big(\cos{n\pi y}-e^{-\frac{(1+i)y}{k\delta}}-e^{-\frac{(1+i)(1-y)}{k\delta}}\big)e^{it}
	\end{multline}
	The vertical velocity from the wall boundary layer $(\frac{\partial\psi_{bl}}{\partial x}\big|_{\eta\rightarrow\infty})$ is neglected because it is of $O(\delta)$. It does not affect the composite solution as the vertical velocity that comes from the bulk at $y=0,1$ is trivial. Using this solution in the leading order kinematic boundary condition (\ref{bdry:kinematic}) helps in determining the deformation $h(y)$ of the interface.
	\begin{flalign}
		h(y) = \sum_{n=0}^{\infty}A_n\bigg(\cos{n\pi y}-e^{-\frac{(1+i)y}{k\delta}}-e^{-\frac{(1+i)(1-y)}{k\delta}}\bigg)e^{it}
	\end{flalign}
	The assumption of pinned interface implies, $h(y)$ must be zero at $y=0,1$. This means $n$ must be even. We incorporate this by transforming $n\rightarrow2n$. Hence the leading order velocity field and interface deformation are given as,
	\begin{multline}
		u_0=\sum_{n=0}^{\infty}A_n\big(e^{-2kn\pi s}+4i\delta^2k^2n^2\pi^2(e^{-\xi_ns}-e^{-kn\pi s})\big)\\
		\big(\big(\cos{2n\pi y}-e^{-\frac{(1+i)y}{k\delta}}-e^{-\frac{(1+i)(1-y)}{k\delta}}\big)\enspace\mathbf{i}+\sin{2n\pi y}\enspace\mathbf{j}\big)e^{it}
	\end{multline}
	\begin{equation}
		h(y)=\sum_{n=0}^{\infty}A_n\big(\cos{2n\pi y}-e^{-\frac{(1+i)y}{k\delta}}-e^{-\frac{(1+i)(1-y)}{k\delta}}\big)e^{it}\label{eqn:interfacedeformation}
	\end{equation}
	\subsubsection{Finding the mode amplitudes of interface deformation}
	Towards our objective of finding the resonance frequency of the system, we first determine the amplitudes of each mode of interface oscillation $(A_n)$. For this, we employ the normal stress balance at the interface $x = l(t,y)$. Using domain perturbation we obtain,
	\begin{flalign}
		p_i^{(0)}-p_0+\frac{1}{Re}\tau_{xx}^{(0)}-\frac{1}{\varepsilon We}\kappa\bigg|_{s=0}=0\label{bdry:normalstresspert}
	\end{flalign}
	Substituting the expressions for $p_i^{(0)}$ and $p_0$ in (\ref{bdry:normalstresspert}) and writing them in terms of Fourier components and collecting all the terms with $y$ dependency, results in:
	\begin{multline}
		\bigg(-\frac{\lambda_f^2}{2n\pi k}+6n\pi ik\lambda_fG+4n^2\pi^2\bigg)A_n-\\
		i\lambda_fG\bigg(\frac{4\alpha\big(1-e^{-\frac{\alpha}{k}}\big)}{4k^2n^2\pi^2+\alpha^2}\bigg)\sum_{m=1}^{\infty}A_m(2m\pi)+\frac{\alpha^2}{k^2}\bigg(\frac{4\alpha \big(1-e^{-\frac{\alpha}{k}}\big)}{4k^2n^2\pi^2+\alpha^2}\bigg)\sum_{m=0}^{\infty}A_m=0\label{normalstressbig}
	\end{multline}
	More details of this equation can be found in  Supplementary section 2. Here, $\lambda_f=\omega/\Omega$. Where $\Omega=\sqrt{\Gamma/(\rho a^3)}$ is used for non – dimensionalising the angular frequency. Also $G=\frac{\mu}{\sqrt{\rho\alpha\gamma}}$ is the Ohnesorge number defined as the ratio of internal viscous dissipation to surface energy. A larger $G$ implies a higher overall stability of slug oscillation.  Larger the $G$, lower will be the amplitude $A_0$ for a given driving frequency. Equation \ref{normalstressbig} gives an infinite system of equations going from $n = 1$ to $\infty$. For computational purposes we consider $n = 1$ to $N$. For each equation, the summation over index $m$ is taken upto $N$.\\
	
	The stream function obtained from boundary layer $(\psi_{bl})$ must satisfy the no stress condition (\ref{bdry:tangential}) at the contact line. This yields
	\begin{equation}
		\sum_{n=0}^NA_n=O(\delta^2)=K\delta^2\label{bdry:pinning}
	\end{equation}
	Detailed derivation of this can be found in Supplementary section 3. Here, the constants $A_n$ are determined for a fixed set of $N + 1$ modes. $N$ stands for the maximum number of eigenmodes we use in the computation. We use a simple matrix inversion to solve the $N+1$ equations (\ref{normalstressbig}, \ref{bdry:pinning}) and thereby find $A_n$. The magnitude of $A_n$ decreases with increase in value of $n$.\\
	
	$\delta$ is of the order of 0.1 for our system and $K$ is a small positive non – zero constant. The values of $A_n$s are insensitive to value of $K$ in the range 0.01 to 10. We choose $K = 0.1$ for all calculations.\\
	
	We express the interface deformation \ref{eqn:interfacedeformation} in terms of a Fourier cosine series for determining the resonant frequency of various modes to facilitate comparison with experiments.
	\begin{equation}
		h(y)=\sum_{n=0}^Na_ne^{i\phi_n}\cos{2n\pi y}
	\end{equation}
	These $a_n$s indicate the amplitudes of different Fourier modes of interface deformation. See Supplementary section 4 for further details. We obtain the $a_n$ values and the interface deformation for a range of driving frequencies $(\omega)$. The driving frequency at which $a_n$ attains a maximum for each mode (defined by $n$) corresponds to the resonance frequency of that mode.
	\section{Numerical analysis}
	The analytical approach helps determine the resonant frequency for each surface mode. We numerically simulate the system at these frequencies to obtain the interface deformation. This helps us obtain an independent verification of the semi analytical results. For a fair comparison, we use the far field pressure found analytically from the earlier analysis for driving the flow in the numerical simulations (See Figure \ref{Figure 1}).\\
	
	We model the system shown in Figure \ref{Figure 1} using Volume of Fluid approach in ANSYS (ver. 2021 R1). The length of liquid in the channel (10 mm) is chosen much larger than the length of the gas slug to mimic the unbounded nature of the liquid. We impose the pressure that was found theoretically from semi analytical investigation, i.e., $p_0$  at the right boundary of the domain as a pressure – inlet. Air is assumed to be compressible and follow polytropic expansions and contractions. Water is assumed incompressible. Surface tension between the fluids is taken as 0.0725 N/m. The section of the wall exposed to air is assumed to be super hydrophobic, while the wall region wetted by the liquid is assigned as super hydrophilic. This is done to ‘pin’ the interface to the wall. We peform simulations at the resonant frequencies obtained for different mode amplitudes. The other details used in these simulations can be found in Supplementary section 5.
	\section{Results}
	\subsection{Resonant frequency of different modes}
	For all computations, the total number of modes considered is $N = 30$ (Refer to Supplementary section 4).
	\begin{figure}
		\centering
		\includegraphics[width=0.5\linewidth]{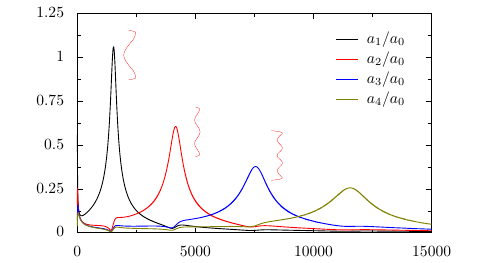}
		\caption{The normalised amplitude modes for a slug of 1.6 mm length and 0.374 mm width with water as surrounding fluid for different driving frequency (in Hz). Maxima for each amplitude modes occur at 1.533 kHz, 4.161 kHz, 7.547 kHz, and 11.545 kHz respectively.}
		\label{Figure 2}
	\end{figure}
	Figure \ref{Figure 2} shows the different normalized Fourier mode amplitudes ($a_n/a_0$) plotted against the driving frequency for a slug of 1.6 mm length and 0.374 mm width surrounded by water. These parameters are chosen to mimic the experiments of \cite{Ryu2010} by using from their experiments, the height of interface from the bottom wall as $a$ and tube diameter as $b$. We see that the maximum amplitude of each succeeding higher mode is lower than the previous one. With an increase in frequency, the different surface modes of oscillation become more prominent. This is observed experimentally in both spherical (\cite{Tho2007}) as well as semi – cylindrical bubbles (\cite{Rallabandi2013}). The predicted resonance frequency of the second mode, 4.161 kHz, is comparable to that obtained experimentally by \cite{Ryu2010}. The experimental images of their work suggest oscillations corresponding to the second mode at $\approx$ 4 kHz.\\
	
\begin{figure}
		\centering
		\begin{tabular}[b]{c}
			\includegraphics[width=.35\linewidth]{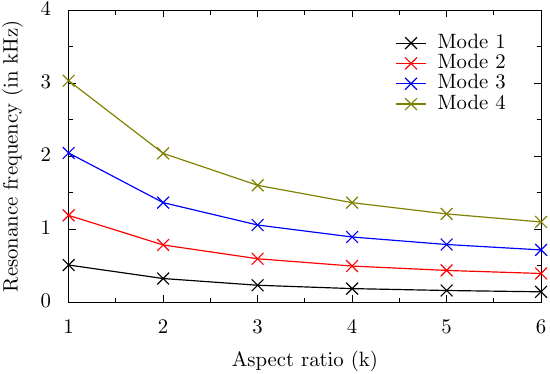} \\
			\small (a)
		\end{tabular} \qquad
		\begin{tabular}[b]{c}
			\includegraphics[width=.35\linewidth]{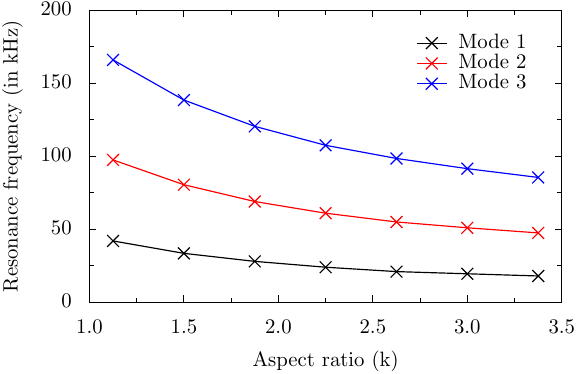} \\
			\small (b)
		\end{tabular}
		\caption{Dependence of resonance frequency  with  aspect ratio for milli (a) and micro (b) channels. Fluid  properties correspond to water at room temperature. The width of the channel ‘$b$’ values are (a) 1.6 mm and (b) 80 $\mu$m.}
		\label{Figure 3}
	\end{figure}

	An important parameter that can be used to alter the resonant frequency of the system is the slug size. The dependence of resonance frequency on the aspect ratio of the slug was investigated analytically and is shown in Figure \ref{Figure 3}. We see a decrease in resonance frequency with increase in aspect ratio of the slug. This trend is similar to that of Minnaert frequency which decreases with increase in size of the bubble. Our model shows that the resonant frequency of a slug in a microchannel is two order of magnitude higher than that in a millichannel.
	\subsection{Interface deformation at resonance frequencies}
\begin{figure}
		\centering
		\begin{tabular}[b]{c}
			\includegraphics[width=.35\linewidth]{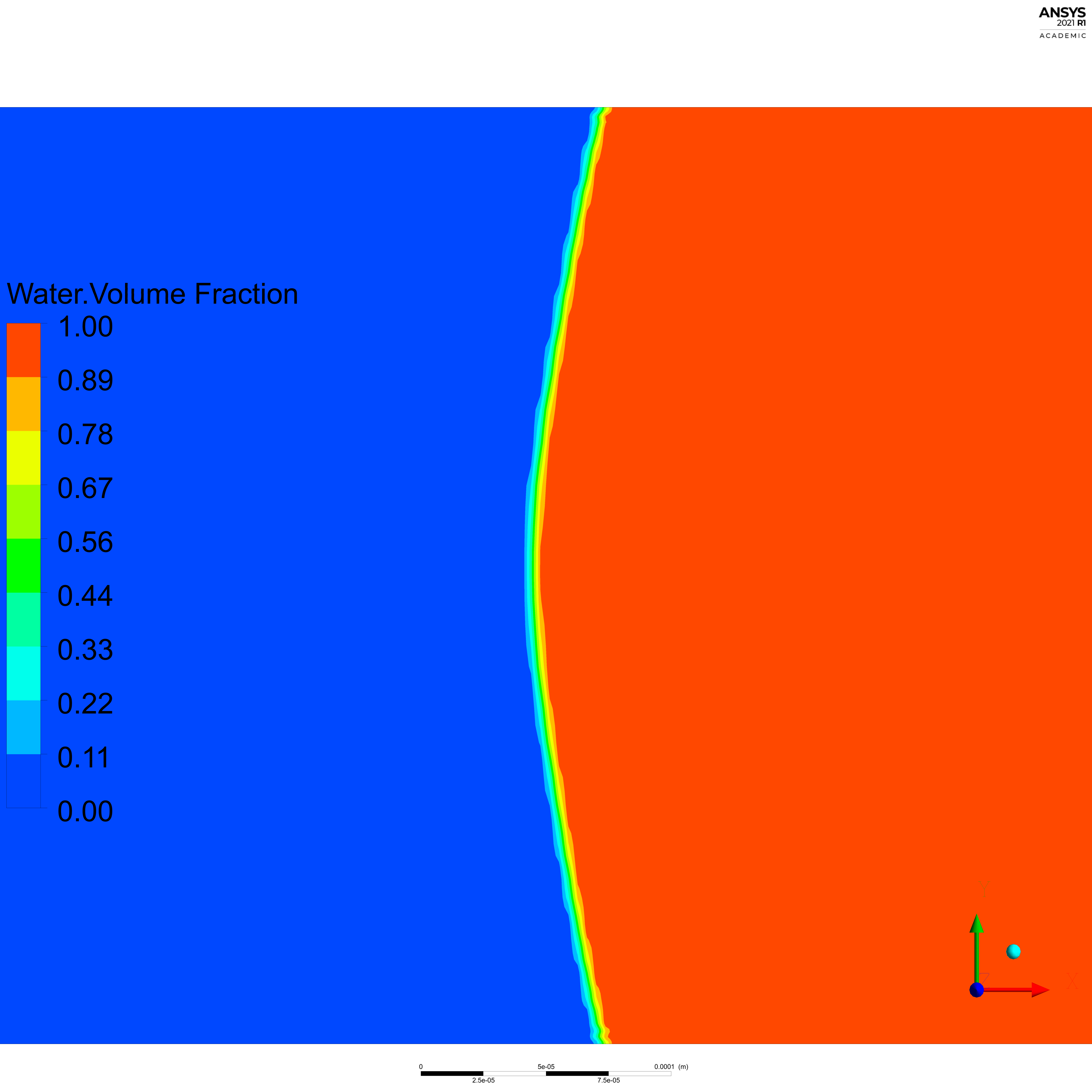} \\
			\small (a) First mode: 1.333 kHz
		\end{tabular} \qquad
		\begin{tabular}[b]{c}
			\includegraphics[width=.35\linewidth]{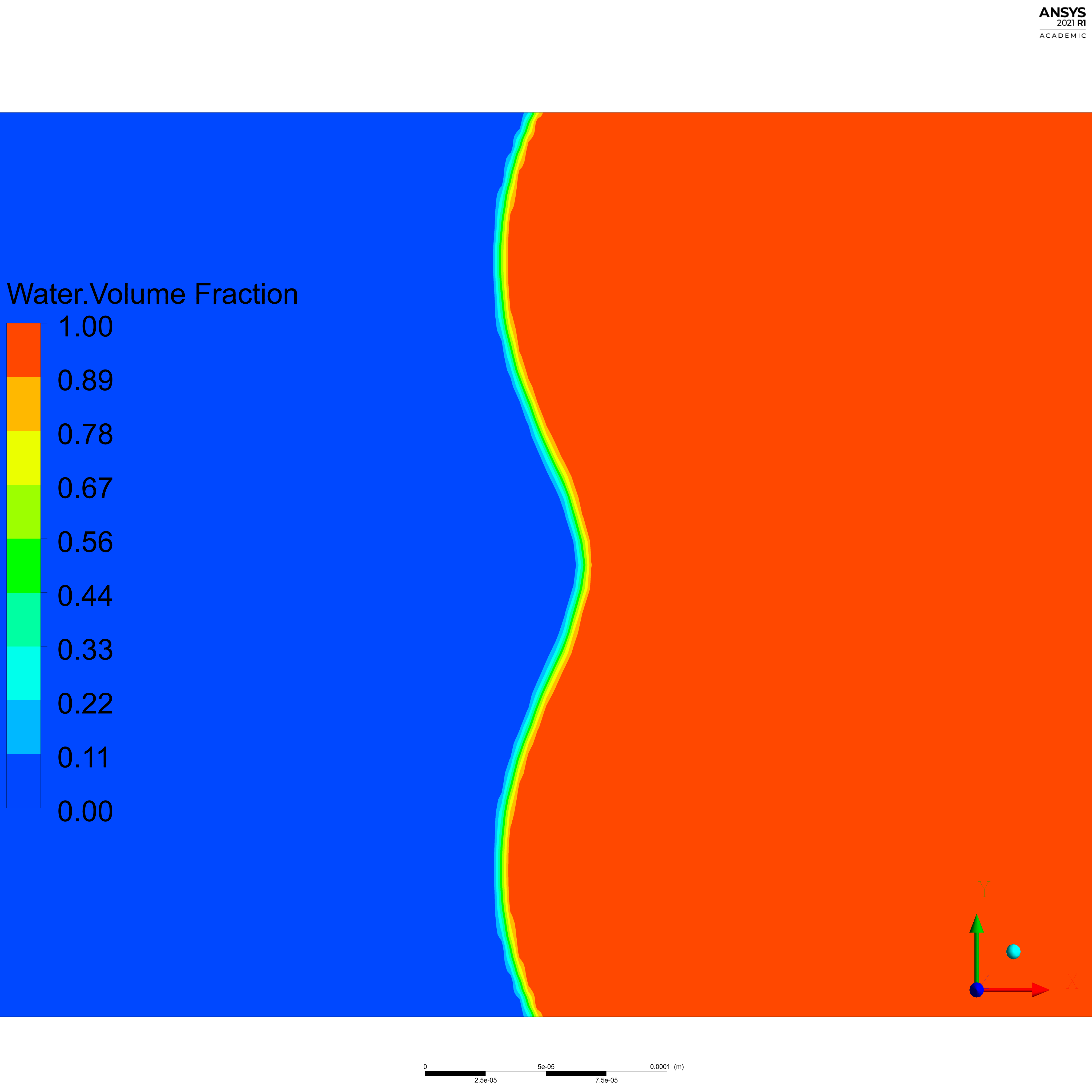} \\
			\small (b) Second mode: 4.161 kHz
		\end{tabular}
		\caption{The interface deformation obtained at two modes through numerical simulations for the first two modes. The length of the slug is 1.6 mm, and the width of the channel is 0.374 mm.}
		\label{Figure 4}
	\end{figure}

	We show the numerically obtained interface deformation from simulations done on ANSYS Fluent at resonant frequencies of the first two modes in Figure \ref{Figure 4}. The numerical simulations confirm the occurrence of higher  modes at increased frequency as predicted in Figure \ref{Figure 2}. We see that for the first mode there is only one extrema, while for the second mode there are three extrema. Hence, there is good qualitative agreement at the resonant frequencies for the first and second modes.\\
	
\begin{figure}
		\centering
		\includegraphics[width=0.45\linewidth]{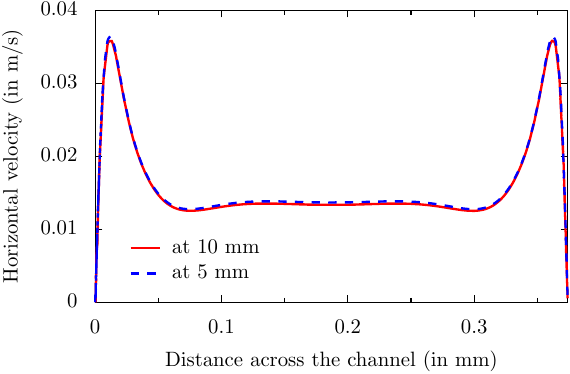}
		\caption{Horizontal leading order velocity as a function of $y$ at two different distances (i) 5 mm (middle of channel) and (ii) 10 mm (outlet) from the left most wall See Figure \ref{Figure 1}.}
		\label{Figure_5}
	\end{figure}

	In Figure \ref{Figure_5}, we show the variation of the horizontal velocity obtained numerically across the width of the channel at two different axial locations $x=5$ and 10 mm (See Figure \ref{Figure 1}). The peaking of velocity near the walls shows that there exists an Eulerian slip at the wall boundary layer as seen in the  semi analytical approach. We infer that this slip is persistent even at long distances from the slug. The distance of the peaks from the wall do not vary much with axial position. This indicates that the wall boundary layer thickness varies negligibly with distance from the slug interface.
	\section{Conclusions}
	The resonance frequency of a bubble pulsating in an unbounded liquid found from potential theory is Minnaert frequency $\big(f_{\text{Minnaert}}=\frac{1}{2\pi a}\big(\frac{3\gamma p_A}{\rho}\big)^{\frac{1}{2}}\big)$. This fails to include viscous effects of the liquid and effect of confinement. In this work we include viscous effects to determine the resonance frequency of gas slugs under confinement. A rectangular gas slug operating at high Strouhal numbers is considered. This allows us to obtain the resonant frequency using a perturbation series approach with the inverse of Strouhal number as the perturbation parameter. The predictions from the semi analytical approach agree with the experiments of \cite{Ryu2010}. These experiments were done in a similar system as that analysed here.\\
	
	We also use ANSYS Fluent to validate the results obtained from our semi analytical approach. The numerical simulations capture the dominant modes at the resonant frequencies predicted by the analytical approach.\\
	
	In summary, we have used a perturbation series solution to identify the resonant frequencies of a gas slug trapped in a milli channel. This study can be used by experimentalists to operate their Lab on Chip devices efficiently.

\bibliographystyle{jfm}
\bibliography{FLMguide}

\end{document}